\documentclass[journal,twoside,web]{ieeecolor}
\usepackage{tmi}
\usepackage{cite}
\usepackage{amsmath,amssymb,amsfonts}
\usepackage{algorithmic}
\usepackage{graphicx}
\usepackage{textcomp}
\usepackage{breqn}
\newcommand{\beginsupplement}{%
	\setcounter{table}{0}
	\renewcommand{\thetable}{S\arabic{table}}%
	\setcounter{figure}{0}
	\renewcommand{\thefigure}{S\arabic{figure}}%
	\setcounter{section}{0}
	\renewcommand{\thesection}{S \arabic{section}}%
}
\DeclareMathOperator{\sign}{sign}
\DeclareMathOperator{\VENC}{VENC}
\DeclareMathOperator{\LT}{\mathcal{L}_{t}}
\DeclareMathOperator{\LM}{\mathcal{L}_{m}^{k}}
\DeclareMathOperator{\LR}{\mathcal{L}_{r}^{k}}
\DeclareMathOperator{\LI}{\mathcal{L}_{I}^{k}}
\DeclareMathOperator{\SUMM}{\sum_{k=1}^{M_{m}}}
\DeclareMathOperator{\SUMMINT}{\sum_{k=1}^{M_{I}}}
\DeclareMathOperator{\SUMMRES}{\sum_{k=1}^{M}}
\DeclareMathOperator{\Mm}{M_{m}}
\DeclareMathOperator{\MI}{M_{I}}
\DeclareMathOperator{\LMS}{\mathcal{L}_{m}}
\DeclareMathOperator{\LIV}{\mathcal{L}_{init}^{V}}
\DeclareMathOperator{\LIA}{\mathcal{L}_{init}^{A}}
\DeclareMathOperator{\LISI}{\mathcal{L}_{init}^{S}}
\DeclareMathOperator{\LBV}{\mathcal{L}_{b}^{V}}
\DeclareMathOperator{\LBA}{\mathcal{L}_{b}^{A}}
\DeclareMathOperator{\LBSB}{\mathcal{L}_{b}^{S}}
\DeclareMathOperator{\LRS}{\mathcal{L}_{r}}
\DeclareMathOperator{\LIS}{\mathcal{L}_{I}}
\DeclareMathOperator{\LMB}{\mathcal{L}_{m}^{b}}
\DeclareMathOperator{\LIAB}{\mathcal{L}_{init}^{A^b}}
\DeclareMathOperator{\LBAB}{\mathcal{L}_{b}^{A^b}}
\DeclareMathOperator{\LMCVS}{\mathcal{L}_{m}^{CVS}}

\DeclareMathOperator{\LPT}{\mathcal{L}^{P}}

\def\BibTeX{{\rm B\kern-.05em{\sc i\kern-.025em b}\kern-.08em
    T\kern-.1667em\lower.7ex\hbox{E}\kern-.125emX}}
\begin{document}
%\bstctlcite{IEEEexample:BSTcontrol}
%\title{Improved cerebral hemodynamics predictions: Combining angiography and transcranial Doppler ultrasound via physics-informed neural networks}
\title{Physics-informed neural networks for improving cerebral hemodynamics predictions}

\author{Mohammad Sarabian, Hessam Babaee, Kaveh Laksari* \thanks{The work was supported by the National Institutes of Health (NIH) National Institute of Neurological Disorders and Stroke (NINDS) grant number R03NS108167.}
\thanks{M. S. is with the Department of Biomedical Engineering, University of Arizona, Tucson, Az, United States (email: msarabian@email.arizona.edu).}
\thanks{H. B. is with the Department of Mechanical Engineering, University of Pittsburgh, Pittsburgh, PA, United States (email: h.babaee@pitt.edu).}
\thanks{K. L. is with the Department of Biomedical Engineering and Department of Aerospace and Mechanical Engineering, University of Arizona, Tucson, Az, United States (email: klaksari@arizona.edu).}}

\maketitle

\begin{abstract}
Determining brain hemodynamics plays a critical role in the diagnosis and treatment of various cerebrovascular diseases. %Therefore, there is a significant need to provide rapid, reliable, and physiologically accurate hemodynamic data with high spatiotemporal resolution for intracranial arteries. 
In this work, we put forth a physics-informed deep learning framework that augments sparse clinical measurements with fast computational fluid dynamics (CFD) simulations to generate physically consistent and high spatiotemporal resolution of brain hemodynamic parameters. Transcranial Doppler (TCD) ultrasound is one of the most common techniques in the current clinical workflow that enables noninvasive and instantaneous evaluation of blood flow velocity within the cerebral arteries. However, it is spatially limited to only a handful of locations across the cerebrovasculature due to the constrained accessibility through the skull's acoustic windows. Our deep learning framework employs \textit{in-vivo} real-time TCD velocity measurements at several locations in the brain and the baseline vessel cross-sectional areas acquired from 3D angiography images, and provides high-resolution maps of velocity, area, and pressure in the entire vasculature. We validated the predictions of our model against \textit{in vivo} velocity measurements obtained via 4D flow MRI scans. %, which provides time-dependent three-dimensional blood velocity in the entire brain, albeit with low spatiotemporal resolution. 
We then showcased the clinical significance of this technique in diagnosing the cerebral vasospasm (CVS) by successfully predicting the changes in vasospastic local vessel diameters based on corresponding sparse velocities measurements.  
The key finding here is that the combined effects of uncertainties in outlet boundary condition subscription and modeling physics deficiencies render the conventional purely physics-based computational models unsuccessful in recovering accurate brain hemodynamics. Nonetheless, fusing these models with clinical measurements through a data-driven approach ameliorates predictions of brain hemodynamic variables. 
%We show this capability by generating synthetic data through performing one-dimensional CFD modeling of pulsatile blood flow within the brain arterial tree before and after vessel spasm. 
\end{abstract}


\section{Introduction}
\label{sec:introduction}
\IEEEPARstart{A}{dvances} in medical imaging technologies have brought about quantifiable improvements in clinical diagnosis, patient management, and clinical outcomes \cite{Wintermark2008}. 
Novel imaging modalities are integral to our pathophysiological knowledge from the molecular level to organ systems, and have led to excitement about possibilities of personalized and predictive medicine \cite{winslow2012}. However, solely relying on  medical imaging has manifested significant limitations \cite{Bredno2010,Flint2010,flint2013} since most diseases are characterized by complex and dynamic interactions that require high spatiotemporal resolution, which is a noted advantage of computational models. 
On the other hand, although computational modeling has made significant progress in providing a quantitative understanding of complex biological processes and their function in disease, their clinical adoption requires sufficient evidence  to convince regulatory agencies to approve clinical trials. This has led to a search towards ``the perfect model'', which has proven elusive so far, and might remain so for years to come, due to the simplifying assumptions and physics deficiencies inherent in modeling. 
A natural question then arises: Can the current modeling approaches, despite all their shortcomings, provide useful information to augment the clinical measurements or should we do away with imperfect computational models for the time being? The fundamental challenge in the task of combining computational models and clinical measurements is the varying accuracy and uncertainty in each source of prediction. 
In this work, we utilize recent advances in the field of deep learning to go beyond the single-source clinical diagnosis to a multi-source paradigm, where computational simulations and clinical measurements are assimilated with the goal of rectifying the paucity of clinical measurements by utilizing physics-based models. This methodology could enable lab-to-bedside deployment of a vast array of existing and future computational models and ultimately lead to a paradigm shift in clinical workflow. 
\par 
We will develop our framework for cerebral blood flow hemodynamics, which constitutes a significant health marker for clinical diagnoses. As an example for other clinical applications, we will test our approach for CVS, which afflicts 70\% of subarachnoid hemorrhage (SAH) patients. SAH is a type of stroke typically due to intracranial aneurysm rupture with an annual incidence of 8-10 per 100,000 adults \cite{linn1996incidence,pluta2009cerebral,melis2019improved}. CVS, defined as focal or diffuse transient narrowing of the large intracranial arteries, typically occurs between day 4 to 14 after hemorrhage and accounts for neurologic deterioration in a third of SAH survivors. To improve the likelihood of early detection of CVS, the patients stay in the intensive care unit (ICU) for a period of two weeks and undergo frequent neurologic, physiologic, and neuroimaging examinations. Computed tomography angiography (CTA) is commonly obtained at presentation of the patient to accurately assess their vascular architecture and screen for cerebral aneurysms. However, high irradiation dose and administration of exogenous contrast agents prompt health concerns and limit the application of repeat CTA studies. Instead, TCD ultrasonography is performed at bedside on a daily basis during the CVS risk window. Despite its ease of use and while it provides reasonable temporal resolution, applicability of TCD as a stand-alone neuromonitoring tool is limited, since it can only provide velocity time series at a limited number of location across the Circle of Willis (CoW), accessible through the skull's acoustic windows. Furthermore, contrary to CTA, which gives geometrical features and branching patterns of the entire cerebrovasculature, TCD provides quantitative blood flow velocity measurements at several locations in the vasculature. Changes in blood flow velocities measured through TCD are used as a surrogate for change in vessel lumen diameters. 
However, in terms of the flow hemodynamics, elevated mean flow velocities are not always linked only to vessel narrowing, and could for example be due to systemic changes, such as high blood pressure, which might lead to higher false-positive decisions \cite{ryu2017numerical}.
As a result, the sensitivity and specificity of TCD for diagnosing CVS, while reasonable, are significantly lower than CTA (sensitivity: 67\% vs. 75.6\%, and specificity: 69\% vs 95.3\%, for TCD and CTA, respectively \cite{geraghty2017delayed,lysakowski2001transcranial,kumar2016vasospasm}). 
There is a notable need to find an alternative and effective tool for diagnosing this devastating disease that is as accurate as CTA scans without its limitations.
Challenges in reliably measuring cerebrovascular structure and hemodynamic properties in the clinic, have motivated the use of CFD modeling. However, while computing power and numerical schemes have significantly improved and resulted in high-resolution, subject-specific blood flow simulations of complex vascular topologies \cite{liu2020state}, clinical adoption of such models has been hampered by the lack of rigorous clinical validation, high computational cost, tedious mesh generation, parameter calibration, and most importantly uncertainty in boundary condition subscriptions. Recent advances in the field of deep learning have overcome the challenge of combining two separate datasets with different levels of accuracy and uncertainty \cite{BPCK16,raissi2019physics,raissi2020hidden,meng2020composite,kissas2020machine,BBDCK20}, which can lead to a  paradigm shift in quantitative approaches towards clinical diagnosis, wherein by augmenting the highly accurate (yet sparse) clinical measurements with the more-readily available computational models in order to increase the spatiotemporal resolution while maintaining the same level of accuracy of the clinical measurements. 
In particular, physics-informed neural networks (PINN) can resolve the above pitfalls which has plagued the pure physics-based CFD models and improve the prediction of hemodynamic variables significantly by combining the clinical measurements with the physics-based models \cite{raissi2020hidden,kissas2020machine,RBG19,MBK21}. Recently, this method was implemented for a rather simple geometry at the carotid bifurcation and trained using \textit{in vivo} flow velocity and wall displacement measurements obtained via 4D flow MRI and segmenting 2D cine images, respectively \cite{kissas2020machine}. 
4D flow MRI is a non-invasive technique that provides time-dependent, three-dimensional blood velocity measures within the vasculature, from which a comprehensive representation of the blood flow hemodynamics could be extracted \cite{markl20124d,rivera20164d,azarine2019four}. However, this technique is not common in the current clinical settings and is mainly limited to fundamental research endeavors as scan times can become prohibitively long, especially when the high spatial resolution or large volumetric coverage are required \cite{markl20124d,azarine2019four}. 
% However, neither 4D flow MRI or cine imaging are commonly acquired in the current clinical settings. 
Furthermore, the wall displacements of intracranial arteries cannot be accurately measured due to the small variation of brain lumen diameters in a cardiac cycle and limited spatial resolution of the MRI data \cite{alastruey2007modelling,park2019efficient}. 
As a result, extending this method to the brain vasculature while strictly using clinically available measurements poses significant challenges given the highly more complex cerebrovascular architecture, including the number of branches and significantly more tortuous nature of the vessels, compared to the carotid bifurcation.  
% }

\begin{figure}[htbp!]
	\centering
	\includegraphics[trim={0cm 0cm 0cm 0cm},width=1\linewidth]{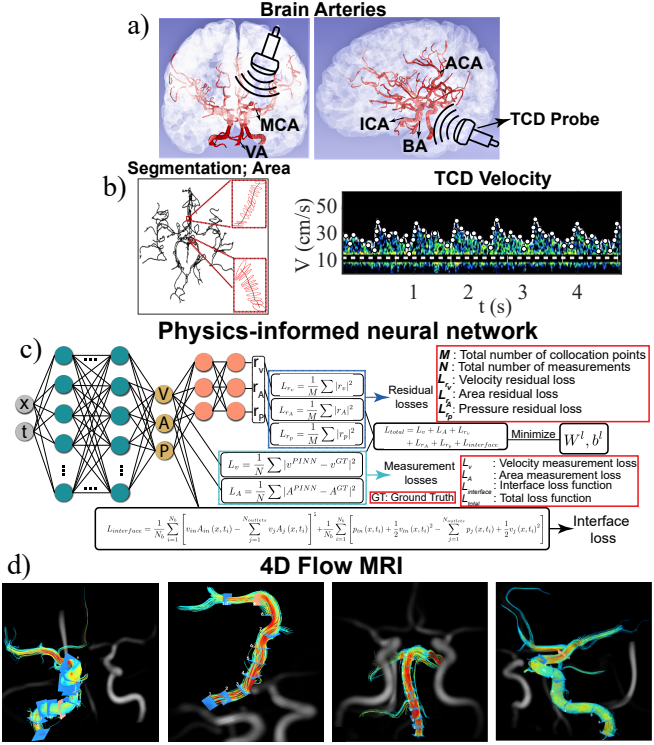}
	\vspace{-4ex}
	\caption{Procedure to generate and validate the cerebral hemodynamics in the CoW: \textbf{a)} 3D rendering of brain vasculature with several major arteries; MCA: middle cerebral artery, VA: vertebral artery, BA: basilar artery, ICA: internal carotid artery, ACA: anterior cerebral artery. \textbf{b)} Vessel segmentation of brain vasculature to extract branching patterns and vessel cross-sectional areas \cite{deshpande2021automatic} in addition to the TCD velocity time-series. \textbf{c)} The structure of the PINN model taking the velocities from TCD and the baseline areas from segmentation, and generating brain hemodynamics in the arterial tree. \textbf{d)} Representative example of 3D blood flow streamlines resulting from 4D flow MRI that we employ to validate the PINN predictions. From left to right, the streamlines are shown for left ICA and MCA, A1 and A2-segments of left ACA, BA, and right ICA, MCA, and ACA at peak systolic, respectively.}
	\label{fig::fig1}
\end{figure}

% \textcolor{blue}{
In the current study, we first present a deep learning framework that can predict blood flow hemodynamics in the entire brain vasculature by utilizing subject-specific 3D angiography and TCD measurements, and validate our modeling predictions against 4D flow MRI measurements (\textbf{Fig. \ref{fig::fig1}}). We extend the current state of the art by designing the neural network, called ``area surrogate physics-informed neural network (ASPINN)", without relying on vessel wall displacements or spatially-resolved 4D flow MRI data. Once trained, the model can predict local blood flow velocity, vessel cross-sectional area, and blood pressure across the entire brain vasculature in a matter of seconds. We further demonstrate the clinical efficacy of our approach in diagnosing CVS, based on synthetically simulated blood flow data within healthy baseline and spastic CoW vasculature.\\

\section{Methods} \label{sec::Methods}
\subsection{Physics-informed neural networks} \label{sec: PINN}
%Here we briefly present the basics of the physics-informed neural network framework and the reader is referred to \cite{kissas2020machine} for further details.
%\par 
%In brief, 
In this framework, the solution of partial differential equations is parametrized by a neural network that is trained to fit the clinical measurements, while being constrained to follow physical laws. Specifically, one neural network (NN) is trained for each CoW arterial segment to generate unknown blood flow variables, \textit{i.e.}, blood velocity, wall displacement, and blood pressure, such that they satisfy three constraints simultaneously \cite{kissas2020machine}. First, the NNs are trained to fit the measured clinical data. %In this work, the clinical data include velocity time-series at few cross-sections obtained from TCD ultrasound, and equilibrium vessel cross-sectional areas (at diastolic phase) in the entire vasculature obtained from segmentation of 3D-ToF scan. In order to train neural networks properly, we also require wall displacements at few cross-sections that could be the same as the cross-sections that we have access to velocity data. Nonetheless, as mentioned previously, due to the lack of 2D cine images for the brain arteries and the limited spatial resolution of these scans, we compute the cross-sectional area time-series by developing the ASPINN model (explained in section \ref{sec:Modifications of physics-informed neural network}). 
Second, they are penalized to satisfy the underlying physical conservation laws governed by a reduced-order one-dimensional (1D) model of pulsatile blood flow \cite{sherwin2003one}. Third, they are trained to satisfy the conservation of mass and momentum across all interface points in the arterial tree. 
% \par
The total loss function is a sum of three individual contributions: measurement loss (corresponds to fitting the clinical measurements), residual loss (to penalize the 1D equations not being satisfied on random points in the space-time domain, known as the ``collocation points"), and interface loss (to satisfy the conservation of mass and momentum on the junctions and bifurcations) (see \textbf{Fig. \ref{fig::fig1}c}). The mathematical expression of the total loss function is as follows:

\begin{equation} \label{eq:loss_total}
\LT = \SUMM \LM + \SUMMRES \LR + \SUMMINT \LI,
\end{equation}
where $\LT$ is the total loss function, %that should be minimized by the neural network,
$\LM$ is the measurement loss of vessel $k$ at which %in the arterial network 
we have collected the clinical data, $\LR$ is the residual loss of $k^{th}$ vessel,  $\LI$ is the interface loss function at the $k^{th}$ interface point, %in the CoW arterial tree,
M is the total number of vessels in the arterial network (M = 23), $\Mm$ is total number of vessels in which we have measurements ($\Mm = 9$), and $\MI$ is the total number of interface points ($\Mm = 11$). %in the network which is 11 in this work.
\par 
The measurement loss is computed via: $\LMS = \LIV + \LIA + \LBV + \LBA,$ 
%for each vessel that we have clinical data is computed via:
%\begin{equation} \label{eq:loss_measuremenst}
 %   \LMS = \LIV + \LIA + \LBV + \LBA, 
%\end{equation}
where $\LIV$, $\LBV$, $\LIA$, and $\LBA$ denote the measurement losses corresponding to the initial and boundary conditions of velocity and area, respectively. 
%are the measurement losses corresponding to the initial conditions for velocity and area respectively, and $\LBV$, $\LBA$ are the measurement losses corresponding to the boundary conditions for velocity and area respectively. 
Note that here by ``boundary conditions" we mean the spatial positions in the CoW where we have access to TCD data and these points need not necessarily be located on the boundaries of the network, which is one of the main advantages of the PINN approach as compared to conventional pure CFD models. 
%For instance, in our case, none of the velocity measurement points are placed at the boundaries of the arterial tree (see \textbf{Fig. \ref{fig::fig2}a}). Indeed, this is one of the advantages of the data-driven approach as compared to conventional pure CFD models where their outputs strongly depend on the subscription of physiologically correct inlet and outlet boundary conditions. 
%The spatial positions of TCD velocity time-series data (training data) in each vessel, shown as blue cross markers in \textbf{Fig. \ref{fig::fig2}a}, are specified via comparison against the 4D flow MRI data.
To compute the positions of TCD data in each vessel, first, we compute the velocity at equilibrium position (diastolic phase) in each segment of the CoW by integrating the mass conservation law, \textit{i.e.}, equation \ref{eq:continuity} below, using the TCD velocity and ToF vessel area data. Then, we compare the computed equilibrium velocity against the same quantity obtained from 4D flow MRI. %Consequently, the precise spatial positions of the velocity training data (TCD data) are determined. 
\par 
The initial and boundary measurement losses are the mean square error (MSE) loss functions and defined as:
\begin{equation} \label{eq:loss_measurement_init_veloci_area}
\LISI = \frac{1}{N_{x}} \sum_{i=1}^{N_{x}} \left(  |S^{*}\left(x_{i},0\right) - S\left(x_{i},0\right)|^{2}       \right),  
\end{equation}

\begin{equation} \label{eq:loss_measurement_bound_velocity_area}
\LBSB = \frac{1}{N_{t}} \sum_{i=1}^{N_{t}} \left(  |S^{*}\left(x_{0},t_{i}\right) - S\left(x_{0},t_{i}\right)|^{2}       \right),  
\end{equation}
%\begin{equation} \label{eq:loss_measurement_init_velocity}
%\LIV = \frac{1}{N_{x}} \sum_{i=1}^{N_{x}} \left(  |V^{*}\left(x_{i},0\right) - V\left(x_{i},0\right)|^{2}       \right),  
%\end{equation}
%\begin{equation} \label{eq:loss_measurement_init_area}
%\LIA = \frac{1}{N_{x}} \sum_{i=1}^{N_{x}} \left(  |A^{*}\left(x_{i},0\right) - A\left(x_{i},0\right)|^{2}       \right),  
%\end{equation}
%\begin{equation} \label{eq:loss_measurement_bound_velocity}
%\LBV = \frac{1}{N_{t}} \sum_{i=1}^{N_{t}} \left(  |V^{*}\left(x_{0},t_{i}\right) - V\left(x_{0},t_{i}\right)|^{2}       \right),  
%\end{equation}
%\begin{equation} \label{eq:loss_measurement_bound_area}
%\LBA = \frac{1}{N_{t}} \sum_{i=1}^{N_{t}} \left(  |A^{*}\left(x_{0},t_{i}\right) - A\left(x_{0},t_{i}\right)|^{2}       \right),  
%\end{equation}
where $N_{x}$ and $N_{t}$ denote the total number of spatial and temporal measurement points, $S$ and $S^{*}$ represent measurement and predicted blood velocities $V$ or vessel cross-sectional areas $A$ respectively. Finally, $x_{i}$, $t_{i}$ represent the spatial and temporal positions, and $x_{0}$ is the spatial position in the vessel at which we have collected clinical data. 
%Note that we can compute the area initial loss $\LIA$ from segmenting the ToF scan and also velocity boundary loss $\LBV$ from the TCD ultrasound velocity data. 
To compute area boundary loss $\LBA$, we require area time-series and due to the lack of clinical data, we have developed an ASPINN framework to recover this information, \textit{i.e.}, $A\left(x_{0},t_{i}\right)$ (see section \ref{sec:Modifications of physics-informed neural network}). %In addition, the velocity initial loss $\LIV$ for each vessel is computed by integrating the continuity equation, \textit{i.e.,}  equation \ref{eq:continuity}. 
Further, we smoothed out the velocity waveforms employing Gaussian Processes (GP) regression \cite{rasmussen2003gaussian} with a periodic kernel function before feeding them to the PINNs framework.   
%===============================================================
\par 
The second contribution to the total loss function is residual loss $\LRS$. %This contribution corresponds to the collocation points randomly distributed inside the arterial domains ($x-t$ plane). 
We have selected 2000 points using the Latin-hypercube sampling strategy \cite{stein1987large}. Over these points, the physics of pulsatile blood flow is enforced by satisfying the governing equations of a one-dimensional model of blood flow dynamics coupled to a pressure-area relationship characterizing the elasticity of the vessel wall. %to describe the transient blood flow through a vessel segment. 
These equations take the following form \cite{formaggia2010cardiovascular,sherwin2003one}:
 
\begin{equation}\label{eq:continuity}
\frac{\partial A}{\partial t} + \frac{\partial Q}{\partial x} = 0,
\end{equation}

\begin{equation} \label{eq:momentum}
\frac{\partial Q}{\partial t} + \frac{\partial}{\partial x} \left(\alpha \frac{Q^{2}}{A}\right) + \frac{A}{\rho}\frac{\partial P}{\partial x} = \frac{f}{\rho},
\end{equation}

\begin{equation} \label{eq:elastic}
P = P_{ext} + \frac{\beta}{A_{0}\left(x\right)} \left(\sqrt{A} - \sqrt{A_{0}\left(x\right)}\right),
\end{equation}
where $Q(x, t)$ is the mass flux, $P(x, t)$ is the average internal pressure over the cross section, $\alpha$ is a non-dimensional momentum flux correction factor, and $\rho$ is the constant mass density of blood. Here, $f$ represents the friction force per unit length and is given by $f=-22\mu \pi V$, where $\mu$ is is the constant blood dynamic viscosity and it provides a good compromise fit to experimental findings \cite{smith2002anatomically}. Here, we set $\alpha=1$ in the convective inertia term of equation (\ref{eq:momentum}) \cite{brook1999numerical}. In equation \ref{eq:elastic}, $\beta = \frac{4}{3} \sqrt{\pi} Eh$, where $h$ and $E$ denote elastic arterial wall thickness and Young’s modulus respectively, with the values adopted from literature \cite{alastruey2007modelling}. Each cross section deforms axisymmetrically independently of the others from a reference state $\left(P,A\right)=\left(P_{ext},A_{0}\right)$, with an external pressure $P_{ext}$ and diastolic cross-sectional area $A_{0}$. Here we assume $P_{ext}=0$ when $A=A_{0}$.  
\par 
Since we enforce equations \ref{eq:continuity}--\ref{eq:elastic} over all collocation points, the corresponding residuals are defined as: $r_{A} := \frac{\partial A}{\partial t} + \frac{\partial Q}{\partial x},$ $r_{v} := \frac{\partial Q}{\partial t} + \frac{\partial}{\partial x} \left(\alpha \frac{Q^{2}}{A}\right) + \frac{A}{\rho}\frac{\partial P}{\partial x} - \frac{f}{\rho},$ $r_{p} := P - P_{ext} - \frac{\beta}{A_{0}\left(x\right)} \left(\sqrt{A} - \sqrt{A_{0}\left(x\right)}\right),$
%\begin{equation}\label{eq:res_continuity}
%r_{A} := \frac{\partial A}{\partial t} + \frac{\partial Q}{\partial x},
%\end{equation}
%\begin{equation} \label{eq:res_momentum}
%r_{v} := \frac{\partial Q}{\partial t} + \frac{\partial}{\partial x} \left(\alpha \frac{Q^{2}}{A}\right) + \frac{A}{\rho}\frac{\partial P}{\partial x} - \frac{f}{\rho},
%\end{equation}
%\begin{equation} \label{eq:res_pressure}
%r_{p} := P - P_{ext} - \frac{\beta}{A_{0}\left(x\right)} \left(\sqrt{A} - \sqrt{A_{0}\left(x\right)}\right),
%\end{equation}
where $r_{A}$, $r_{v}$, and $r_{p}$ represent mass, momentum, and pressure residuals, respectively, and are functions of spatial $x$ and temporal $t$ positions. The partial derivatives in the residual expression are computed using automatic differentiation  \cite{abadi2016tensorflow}. These residuals enforce the neural networks to predict velocity, area, and pressure so as to satisfy the underlying differential equations \ref{eq:continuity}--\ref{eq:elastic}. Eventually, the residual loss function for each vessel in the CoW arterial tree is computed via:

\begin{equation} \label{eq:residual_loss}
\LRS = \frac{1}{N_{f}} \sum_{i=1}^{N_{f}} \left(r_{A}\left(x_{i},t_{i}\right)\right)^{2} +  \left(r_{v}\left(x_{i},t_{i}\right)\right)^{2} + \left(r_{p}\left(x_{i},t_{i}\right)\right)^{2},  
\end{equation}  
where $N_{f}$ is the total number of collocation points in each vessel. 
\par 
The final contribution to the total loss function, \textit{i.e.}, interface loss, corresponds to satisfying the conservation of mass and momentum across all interface points and is computed at each bifurcation point via:

%\begin{equation} \label{eq:loss_interface}
%\begin{split}
%Loss_{interface} &= \frac{1}{N_{b}} \sum_{i=1}^{N_{b}} \left[Q_{in}\left(x_{b},t_{i}\right) - \sum_{j=1}^{N_{outlets}} Q_{j} \left(x_{b},t_{i}\right)\right]^{2} \\
%&+ \frac{1}{N_{b}} \sum_{i=1}^{N_{b}} \left[\zeta_{in}\left(x_{b},t_{i}\right) - \sum_{j=1}^{N_{outlets}} \zeta_{j}\left(x_{b},t_{i}\right)\right],
%\end{split}
%\end{equation} 
\begin{dmath} \label{eq:loss_interface}
\LIS = \frac{1}{N_{b}} \sum_{i=1}^{N_{b}} \left[Q_{in}\left(x_{b},t_{i}\right) - \sum_{j=1}^{N_{outlets}} Q_{j} \left(x_{b},t_{i}\right)\right]^{2} + \frac{1}{N_{b}} \sum_{i=1}^{N_{b}} \left[\zeta_{in}\left(x_{b},t_{i}\right) - \sum_{j=1}^{N_{outlets}} \zeta_{j}\left(x_{b},t_{i}\right)\right]^{2},
\end{dmath}
where $N_{b}$ represents the number of collocation points on the interface boundaries, $Q_{in}$ is the inlet volumetric flow rate to each bifurcation, $x_{b}$ is the spatial position of the interface boundaries, $N_{outlets}$ is the number of outlets at each bifurcation, and $\zeta$ is the total momentum which is the sum of average internal pressure $P$ over the cross-section at each bifurcation and the dynamic pressure. The interface loss function ensures the flow information measurements in one vascular domain can be propagated throughout the neighboring domains \cite{kissas2020machine}. %Indeed, including the interface loss into the total loss function provides extra information on the mass and momentum fluxes to neighboring vessels that results in approximating the flow information even for the vessels that we don't have access to any training data. 
%\par 
%We have performed non-dimensionalization of the different physical quantities to overcome the difficulty imposed by the significant difference in magnitude of the parameters which impacts the magnitude of the back-propagated gradients that adjust the neural network parameters during training \cite{kissas2020machine}. Further, we normalized the inputs to the neural network, \textit{i.e,} $x,t$ to have zero mean and unit variance that relieves the issue of vanishing gradients in deep networks and ensures the robustness of the neural network training via back-propagation \cite{glorot2010understanding}.
\par 
Here, each neural network has 7 hidden layers and 100 hidden units per hidden layer, and an exponential linear unit (elu) activation function is chosen for all the hidden layers. Each neural network's weights and biases $\left\{W^{l},b^{l} \right\}$ of layer $l$ are learned by minimizing the total loss function via stochastic gradient descent using Adam optimizer with default settings \cite{kingma2014adam}. The training is performed for a total number of $500,000$ epochs with the learning rate equal to $10^{-3}$. Moreover, to increase the computational efficiency, we train the neural networks by feeding them mini-batches of size 1024 collocation points. The predictions of neural networks are robust with respect to the chosen batch size.  %, to exhibit the training efficiency of the method.
%\textbf{figure \ref{fig::Suppl.Fig4}}, to exhibit the training efficiency of the method.  

\subsection{Area surrogate physics-informed neural networks} \label{sec:Modifications of physics-informed neural network}
As discussed previously, due to the lack of \textit{in vivo} clinical data for the vessel wall displacements in the brain vasculature, we are not able to compute area boundary loss $\LBA$. Hence, in the current study, we have developed an area surrogate physics-informed neural network (ASPINN) model to generate area time-series at the spatial positions, for which we have collected TCD data. %In other words, the objective of the ASPINN model is to predict the vessel cross-sectional area at $x$ and $t$, given the blood flow velocity at those positions. 
Once the vessel wall displacements are determined, we use them along with the velocity data obtained from TCD ultrasound as training data to the PINN model explained in section \ref{sec: PINN} to generate physically consistent outputs in the entire vasculature. 
\par
\begin{figure}[h!]
	\centering
	\includegraphics[trim={0cm 0cm 0cm 0cm},width=1\linewidth]{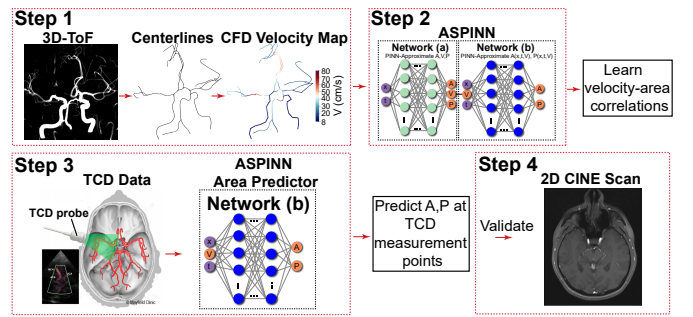}
	\vspace{-4ex}
	\caption{Step-by-step procedure to build the subject-specific ASPINN model for each vessel in the CoW vasculature to generate and validate physics-based area time-series at TCD velocity measurement points.}
	\label{fig::fig5}
\end{figure}

The step-by-step procedure of the ASPINN model is demonstrated in \textbf{Fig. \ref{fig::fig5}}. From the 3D ToF scan, we first performed image segmentation and extract the vessel centerlines using the pathline-based model construction workflow of SimVascular \cite{updegrove2017simvascular}. Then, we perform one-dimensional CFD modeling of blood flow through the vasculature using SimVascular (Step $1$). We set the inlet boundary conditions from the processing of our 4D flow MRI dataset, and for the outlet boundary conditions we use the three-element Windkessel model \cite{westerhof2009arterial}, where we adopted the total arterial compliance and systemic vascular resistance from the literature \cite{alastruey2007modelling}. Next, we build our ASPINN model to learn subject-specific velocity-area correlations for each segment within the CoW network via training two fully-connected, feed-forward PINNs, \textit{i.e.}, Networks (a) and (b) in series (Step $2$). Henceforth, we utilize the ASPINN model (Network (b)) to predict area time series using the \textit{in vivo} TCD velocity time-series $v(t)$ and the spatial position of velocity measurement $x$ as inputs to the neural Network (b) (Step 3). In addition, our ASPINN model generates internal blood pressure time-series averaged over those cross-sections. Finally, we validate the area predictions against segmenting 2D cine scans (Step 4). Note that since the wall displacement of brain arteries cannot be recovered from segmentation of cine images, we have validated our developed ASPINN model predictions against the area time-series resulting from the segmentation of 2D cine scan of aorta/common carotid bifurcation of a healthy subject with the clinical data adopted from \cite{kissas2020machine} (See appendix \ref{sec::appendix_A} \textbf{Fig. S1}).% \textbf{Fig. \ref{fig::Suppl.Fig1}}). 
The velocity-area correlations in each vessel predicted by the optimized ASPINN model along with the comparison against one-dimensional CFD modeling as well as the estimated area time series using the TCD velocity data as input are shown in the appendix \ref{sec::appendix_A} \textbf{Fig. S2}. % \textbf{figure \ref{fig::Suppl.Fig2}}.
\par 
Network (a) of the ASPINN model (step 2 in Fig. \ref{fig::fig5}) is employed to approximate the velocity, area, and pressure at each $x$ and  $t$. Network (b) is designed for approximating the correlation between blood flow velocity and vessel cross-sectional area at each $x$ and $t$. Both Networks (a) and (b) are trained based on the velocity and area information generated by performing one-dimensional CFD modeling of blood flow through the CoW. It is noteworthy to mention that during the training of Network (b) the velocity derivatives with respect to input features, \textit{i.e.}, $\partial{v}/\partial{x}$, $\partial{v}/\partial{t}$ are obtained from Network (a) in each iteration.     
%\begin{figure}[h!]
%	\centering
%	\includegraphics[trim={0cm 0cm 0cm 0cm},width=0.7\linewidth]{./Figures/Figure6} 	
	%	\vspace{-30ex}
%	\caption{A schematic representation of the proposed surrogate model to predict area and pressure time-series at any position in the vasculature given velocity time-series and the spatial position of the velocity data. Network (a) is the original PINN model described in section \ref{sec: PINN} and Network (b) generates area and pressure given velocity.  }
%	\label{fig::fig6}
%\end{figure}
\par 
First, we perform one-dimensional CFD simulation of blood flow to generate velocity, area, and pressure in the entire vasculature. Next, we train Network (a) by employing velocity and area time series at similar spatial positions as TCD measurements (as shown in \textbf{Fig. \ref{fig::fig2}a}) derived from the CFD modeling (see section \ref{sec: PINN}). Note that since we train Network (a) based on the outputs of CFD simulation, we have access to the area boundary conditions and can compute area boundary loss $\LBA$. %The outputs of Network (a) are the high spatiotemporal resolution of blood velocity, area, and pressure. 
Then, we train Network (b) for each vessel in the CoW arterial tree using the outputs of Network (a) to learn subject-specific velocity-area correlations. %Concretely, Network (b) is trained based on the velocity and area time-series at the same spatial positions shown in \textbf{Fig. \ref{fig::fig2}a} as training data. 
The structure of hidden layers of Network (b), the choice of activation function, the learning rate, the type of optimizer, the number of iterations, the mini-batch size, and the number of collocation points are the same as Network (a) and discussed in section \ref{sec: PINN}. 
\par 
Network (b) is trained by minimizing the same total loss function (equation \ref{eq:loss_total}) as Network (a). 
%As previously mentioned, to compute momentum residual $r_{v}$ we need the information of partial derivatives of velocity $\partial v/\partial t$ and $\partial v/\partial x$. We fetch these derivatives from Network (a) in each iteration after Network (a) is fully trained. Nonetheless, the partial derivatives of area and pressure are computed based on the predictions of Network (b). 
The measurement loss for each vessel of Network (b) %in which we provide training data 
is computed via: $\LMB = \LIAB + \LBAB,$
%\begin{equation} \label{eq:loss_measuremenst_Network_b}
%\LMB = \LIAB + \LBAB, 
%\end{equation}
where $\LMB$ is the measurement loss of Network (b), and $\LIAB$ and $\LBAB$ are the measurement losses of Network (b) corresponding to the are initial and boundary conditions respectively, computed as described below:
\begin{equation}
\LIAB = \frac{1}{N_{x}} \sum_{i=1}^{N_{x}} \left(  |A^{*}\left(x_{i},0, V\left(x_{i},0\right)\right) - A\left(x_{i},0,V\left(x_{i},0\right)\right)|^{2}       \right),  
\end{equation}

\begin{equation}
\LBAB = \frac{1}{N_{t}} \sum_{i=1}^{N_{t}} \left(  |A^{*}\left(x_{0},t_{i},V\left(x_{0},t_{i}\right)\right) - A\left(x_{0},t_{i},V\left(x_{0},t_{i}\right)\right)|^{2}       \right).  
\end{equation}
%Note that we provide initial velocity $ V\left(x_{i},0\right)$ and velocity time-series at each spatial position $V\left(x_{0},t_{i}\right)$ based on the outputs of Network (a). Finally, the interface loss is computed via equation \ref{eq:loss_interface}, where the velocity data required at the bifurcation points $x_{b}$ at each iteration are drawn from the predictions of Network (a). 
\par 
Finally, the vessel cross-sectional area time-series predicted by the trained ASPINN model at the spatial positions in the CoW architecture along with the velocity data obtained from TCD ultrasound are served as training data and used to train the PINN model discussed in section \ref{sec: PINN} to generate physically consistent brain hemodynamics in the vasculature.    
%
%
% ==============================================
% ==============================================
\subsection{Cerebral vasospasm diagnosis using PINN} \label{sec:Methods_CVS}
\begin{figure}[h!]
	\centering
	\includegraphics[trim={0cm 0cm 0cm 0cm},width=1\linewidth]{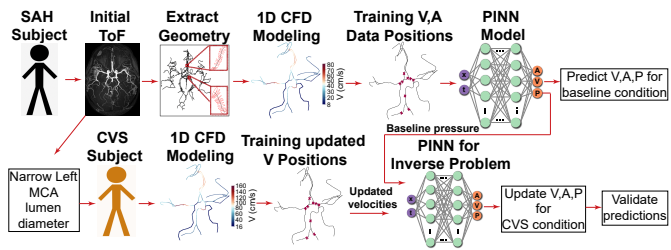} 	\vspace{-4ex}
	\caption{Step-by-step procedure to update the vessel cross-sectional areas and cerebral hemodynamics in the CoW arterial network after CVS employing PINNs for solving an inverse problem.  }
	\label{fig::fig7}
\end{figure}

%The clinical significance of the PINN framework is demonstrated via utilizing this model to diagnose CVS by updating and recomputing the cross-sectional areas (primary variable for CVS diagnosis) of the entire CoW arterial tree. To show this ability of the PINN model, we have generated synthetic training and test data through performing a series of one-dimensional CFD simulations of blood flow through CoW for the baseline condition (no vasospasm) and all the six stages of vessel spasm using SimVascular open-source software \cite{updegrove2017simvascular}.  
%\par
The general overview of our procedure to update vessel cross-sectional areas, and cerebral hemodynamics in the entire vasculature after CVS is illustrated in \textbf{Fig. \ref{fig::fig7}}.
First, we perform image segmentation of the initial angiography scan of a SAH patient to find initial cross-sectional areas, branching pattern, and vessel lengths using SimVascular \cite{updegrove2017simvascular}. Note that as we don't have access to the SAH patient data we develop our model based on the healthy subject and synthetically create a CVS subject as explained below. Next, we generate synthetic data for training and validating the PINNs by performing 1D CFD simulation of the baseline condition (before CVS) using SimVascular \cite{updegrove2017simvascular}. The spatial positions of velocity and area training data are marked as red cross-markers on the centerlines and are selected to be the same as shown in \textbf{Fig. \ref{fig::fig2}a}. We then train a PINN model for each vessel for the baseline condition using the method discussed in section \ref{sec: PINN}. So far, we have estimated the baseline brain hemodynamic features in the vasculature. 
\par 
Next, we use post CVS data to re-evaluate local cross-sectional areas. However, in this study we only synthetically construct CVS cases by shrinking the M1-segment of the left MCA uniformly across the lumen, with no change in vessel shape and perform one-dimensional CFD modeling for all the six CVS cases. Nevertheless, we only use the updated velocity time-series at the same spatial positions as the baseline TCD measurements, along with the pressure distribution predicted by the PINNs at the baseline condition to solve an inverse problem. We retrain the PINN model and update the vessel cross-sectional areas, velocity, and blood pressure in the entire vasculature. Finally, we validate the neural network predictions against the outputs of one-dimensional CFD simulations, which serves as our ground-truth data. Here, we assume that the pressure distributions in the affected vessel do not significantly change compared to the baseline condition (see appendix \ref{sec::appendix_A} \textbf{Fig. S3}).% (see supplementary \textbf{figure \ref{fig::Suppl.Fig3}}).    
\par
The second PINN model (after CVS) is trained by minimizing the same total loss function presented in equation \ref{eq:loss_total}. The residual and interface loss functions remain the same as equations \ref{eq:residual_loss} and \ref{eq:loss_interface}. However, the measurement loss is computed via: $\LMCVS = \LIV + \LBV + \LPT,$
%\begin{equation} \label{eq:loss_measuremenst_CVS}
%\LMCVS = \LIV + \LBV + \LPT, 
%\end{equation}
where $\LMCVS$ is the measurement loss for each vessel after CVS, $\LPT$ denotes the pressure measurement loss and is computed via: $\LPT = \frac{1}{N_{p}} \sum_{i=1}^{N_{p}} \left(  |P^{*}\left(x_{i},t_{i}\right) - P\left(x_{i},t_{i}\right)|^{2}       \right),$
%\begin{equation}
%\LPT = \frac{1}{N_{p}} \sum_{i=1}^{N_{p}} \left(  |P^{*}\left(x_{i},t_{i}\right) - P\left(x_{i},t_{i}\right)|^{2}       \right),  
%\end{equation}
where $N_{p}$ is the total number of pressure training data points in the $x-t$ plane. Here, $P^{*}$ is the predicted pressure by neural networks at the training data points. The second neural network that is trained after CVS has the same architecture as the first one trained before CVS (see section \ref{sec: PINN}), but with more variables to get trained. Concretely, the local equilibrium cross-sectional areas $A_{0}$ of each vessel are now defined as trainable variables in addition to weights and biases of each neural network. 
%$A_{0}$ is either a single variable for a vessel or a vector that is a function of spatial position $x$ of the vessel depending on the original vessel geometry before CVS. 
Thus the set of trainable parameters $\hat{\Theta}$ consists of $\left\{W^{l},b^{l} \right\}_{l=1}^{D}$ and $\left\{A_{0}^{j}\right\}_{j=1}^{N_{x}}$, where $D$ is the number of hidden layers of each neural network. These variables are updated during training by minimizing the total loss function (equation \ref{eq:loss_total}) using the Adam optimizer \cite{kingma2014adam}. 
%\par 
All the algorithms presented in sections \ref{sec: PINN}, \ref{sec:Modifications of physics-informed neural network}, and \ref{sec:Methods_CVS} are implemented in Tensorflow v2.0 \cite{tensorflow2015-whitepaper}, and computations were performed in double precision arithmetic on a single NVIDIA Tesla P200 GPU card. The MRI clinical data acquisition and the pre-and-post processing steps of 4D flow MRI scans, which are based on the procedure presented in \cite{bock2007optimized,markl20124d} are explained in the appendix \ref{sec::appendix_A}.

\section{Results} \label{sec::Results}
\subsection{Brain hemodynamic predictions} \label{sec::Results_hemodynamic}
\begin{figure}[htbp!]
	\centering
	\includegraphics[trim={0cm 0cm 0cm 0cm},width=0.95\linewidth]{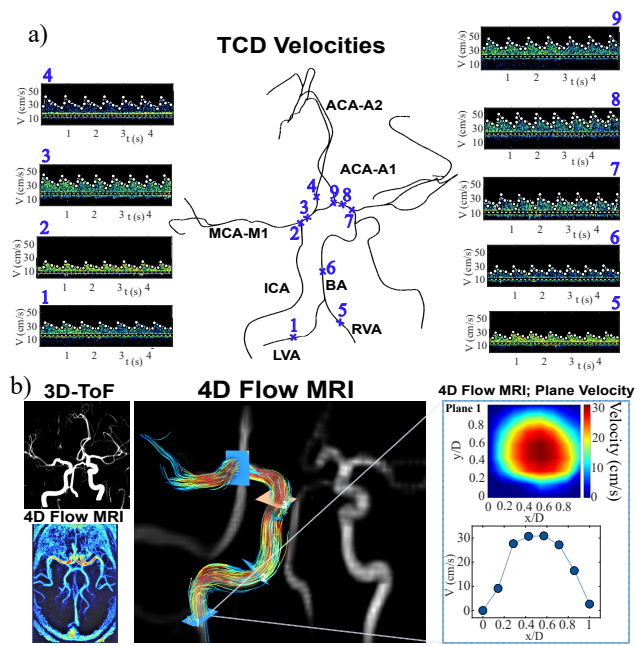} 	%\vspace{-2ex}
	\caption{Clinical data of a healthy human volunteer (30-year old male) used for training and validating of the PINNs. \textbf{a)} TCD ultrasound velocity measurements at nine spatial points shown on the centerline of the CoW arteries. %These velocity measurements are used to train the neural networks to predict the brain hemodynamics and wall displacements in the entire arterial network. 
	\textbf{b)} 3D ToF MRI scan and 4D flow MRI data, representing the CoW architecture (top-left panel) and the three-dimensional peak systolic blood flow distributions (bottom-left panel) in the arterial tree.  
	%The bottom left panel demonstrates the blood flow distribution at peak systolic in the CoW network (thickness of the 2D slab is 4mm). 
	The middle panel shows 3D velocity streamlines at peak systolic through the left ICA and MCA. The dashed blue box represents the cross-sectional blood flow velocity at the plane located at the inlet of the left ICA and the corresponding velocity profile.}
	\label{fig::fig2} 
\end{figure}

We acquired \textit{in vivo} structural properties as well as blood flow velocity data across the CoW of a healthy volunteer as follows. First, using a 2MHz TCD ultrasound probe (DigiLite, Rimed, Inc), we collected flow velocity time-series at nine spatial locations in the CoW, namely, left and right VAs (LVA, RVA), BA, left and right M1-segment of MCA, left and right A1 and A2 segments of ACA (as depicted in \textbf{Fig. \ref{fig::fig2}a}). Comparable to clinical measurements, this data was collected at 14.7 ms temporal resolution and over 40 cardiac cycles. We  use the TCD velocity measurements for training the neural networks as described below. We also performed MRI scans from the same subject, including a ToF angiography scan, which is routinely performed in the clinic and, similar to CTA, is used to build the geometrical features of the CoW for training. 
In addition, we obtained a 4D flow MRI scan, which provides three-dimensional velocity time series maps in the entire CoW vasculature, but is not a part of routine clinical workflow and will only be used for validation purposes (\textbf{Fig. \ref{fig::fig2}b}). The spatial resolution of 4D flow MRI was $1.5\times 1.26\times 2$ $mm^3$, and the temporal resolution is 42 msec over 14 cardiac phases. All the MRI scans were performed on a 3T Siemens Skyra at University of Arizona’s Translational Bioimaging Resource facilities. 
\par 
%Using the ToF angiograms and the TCD velocity time series, we trained a series of PINN for all the vessels belonging to the CoW arterial tree. Each neural network is trained to fit the TCD velocities while being constrained by the physical laws of one-dimensional  model of pulsatile blood flow simultaneously \cite{sherwin2003one}. The inputs to the neural network include local cross-sectional areas, derived from segmentation of the ToF scan \cite{deshpande2021automatic,updegrove2017simvascular}, and the TCD velocity time series at the nine CoW locations mentioned above. Once the parameters of all neural networks are optimized, we can obtain  continuous space- and time-resolved estimates for structural and hemodynamic variables, including blood flow velocity, cross-sectional area, and pressure in the entire CoW tree (see section \ref{sec::Methods} for detailed description of the methodology). 
\par
\begin{figure}[h!]
	\centering
	\includegraphics[trim={0cm 0cm 0cm 0cm},width=1\linewidth]{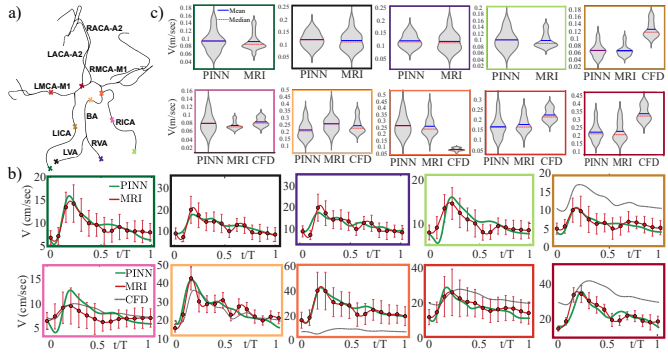} 	
	%	\vspace{-30ex}
	\caption{Comparison of predicted results: \textbf{a)} CoW centrelines, marked by color-coded labels for each of the ten test point. \textbf{b)} Comparison of blood flow velocity time-series between the PINN, 4D flow MRI, and CFD modeling. The error bars in the 4D flow MRI measurements denote the standard deviation of cross-sectional velocities in that particular time frame. \textbf{c)} Velocity distributions in one cardiac cycle at ten test spatial points resulting from PINN, 4D flow MRI, and CFD modeling. The color of each panel corresponds to its spatial location in the arterial tree. The width of the probability density function shows relative distribution density. In panels b) and c), the color of each box corresponds to its spatial location in the arterial tree as marked in a). 
	%Asterisk (*) shows a significant difference between MRI data and CFD predictions of blood flow velocities ($p<0.05$) \textcolor{red}{we should prob remove the * since only 2 of the plots are significant}
	}
	\label{fig::fig3}
\end{figure}

 
We compared blood flow velocity time series predicted by the PINN model at ten different spatial positions in the CoW against the flow velocities measured through 4D flow MRI, which provides more spatial information (\textbf{Fig. \ref{fig::fig3}}). We found good agreement in velocity time evolution between the predictions of neural networks and the \textit{in-vivo} velocity data resulting from 4D flow MRI during each cardiac cycle at every test point considering the uncertainty range of MRI measurements (\textbf{Fig. \ref{fig::fig3}b}). We observed that despite the lack of knowledge about the inlet and outlet boundary conditions, which is a major setback in traditional CFD simulations, and the relatively few velocity training data, our PINN model is capable of closely approximating the physiologically correct blood flow velocities in the entire CoW network. Note that to predict the cerebral hemodynamic parameters using the PINN model, we first utilized our developed ASPINN model to estimate physically-consistent area wave propagation at TCD velocity measurement points, which are shown in \textbf{Fig. S2b} of the appendix \ref{sec::appendix_A} for few locations across the vasculature. Each component of the total loss function of the PINN model, \textit{i.e.}, equation \ref{eq:loss_total}, as a function of the number of training epoch is shown in the appendix \ref{sec::appendix_A} \textbf{Fig. S4}. 
\par 
In order to further demonstrate this advantage, we simulated the blood flow velocities in the same CoW network using a purely physics-based axisymmetric CFD model (SimVascular, Stanford, CA) \cite{updegrove2017simvascular}. We used the physiological velocity values, extracted from the 4D flow MRI data, to prescribe boundary conditions at the four inlets (LICA, RICA, LVA, and RVA). In addition, we used the commonly-used three-element Windkessel model \cite{westerhof2009arterial} for the outlet boundary conditions, and adopted the total arterial compliance and systemic vascular resistance from the literature \cite{alastruey2007modelling}. As can be seen, there are substantial discrepancies between the CFD predictions and the ground-truth MRI measurements. Specifically, although the CFD model can predict the general trend of the blood velocities, it either underestimates or overestimates the velocity range or introduces a time delay in the peak velocity values. This is despite the fact that we used the accurate inlet boundary conditions (from 4D flow MRI measurements), which are normally not available for a subject-specific simulation. This constitutes a significant advantage of our machine learning approach over traditional CFD modeling approaches. 
\par 
In addition, the probability density functions (PDF) of blood velocity over one cardiac cycle resulting from 4D flow MRI measurements, predictions of PINNs, and CFD model outputs are illustrated for a point on each of the CoW branches in \textbf{Fig. \ref{fig::fig3}c}. The distributions highlight the velocity field agreement between the PINNs and the \textit{in-vivo} MRI data and the substantial discrepancies between those and the CFD model. The discrepancies become more apparent as we move further away from the inlets of the CoW arterial network, which indicates the effect of physics-deficiencies and more importantly the uncertainties in the outflow boundary conditions (\textbf{Fig. \ref{fig::fig3}b}). Hence, uncertainties in the outlet boundary condition subscription (due to the lack of knowledge in pressure distributions in the efferent vessels) lead to erroneous predictions of brain hemodynamics by the one-dimensional computational model as can be observed in \textbf{Fig. \ref{fig::fig3}}. Note that the results of the CFD simulations are not shown at the four inlets (LICA, RICA, LVA, RVA) as the flow velocities there are prescribed as inlet boundary conditions directly from the 4D flow MRI measurements. 

% ########################################

% ########################################
\subsection{Cerebral vasospasm diagnosis}
In this section, we utilize our PINN framework developed above to address a specific clinical cerebrovascular disease, \textit{i.e.}, CVS, which requires precise knowledge of brain hemodynamics and vessel cross-sectional areas for diagnosis. As mentioned in the introduction, the primary parameter in diagnosing CVS is the local cross-sectional areas and how they evolve during the length of stay in the ICU compared to the equilibrium cross-sectional areas, extracted from baseline CTA on the day of admission. 
However, high-dose irradiation and administration of contrast agents limit the application of repeat CTA studies. Consequently, in order to approximate updated vessel local diameters, after CVS, instead of CTA, TCD is performed at the bedside daily during the 14-day CVS window, which provides the velocity information at only several spatial positions in the vasculature. Below, we present a data-driven approach to automatically compute and update the local cross-sectional areas in the entire CoW vasculature based on new blood flow velocities obtain from TCD ultrasonography.

\begin{figure}[h!]
	\centering
	\includegraphics[trim={0cm 0cm 0cm 5cm},width=1\linewidth]{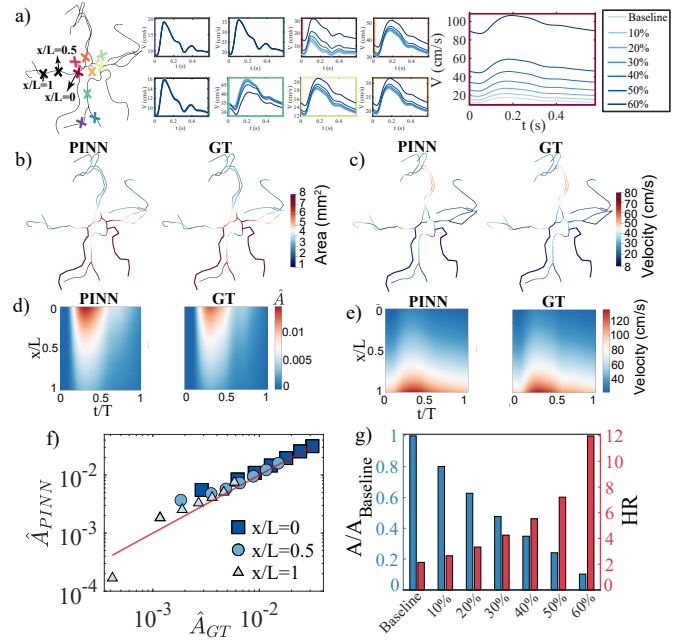} 	
	%	\vspace{-30ex}
	\caption{Application to CVS: \textbf{a)} Location of training data, and their corresponding velocity time-series at baseline and six levels of vasospasm (Box colors correspond to the marker colors on the CoW centerline). The training data are velocity and area for the PINN model that is trained for baseline condition and are velocity and pressure for the PINN model that is trained for CVS cases. \textbf{b-c)} Spatial comparison of 3D systolic area and velocity maps between predictions of PINNs and synthetic ground truth (GT) data in the entire CoW for a CVS case with 30\% uniform reduction in diameter.
	\textbf{d-e)} Spatiotemporal comparison of the area and velocity maps at the left MCA after CVS between PINN predictions and GT data. 
% 	In panels b), c), d), and e) the vessel diameter is uniformly reduced by $30\%$ across the lumen with no change in the vessel shape. 
	\textbf{f)} Non-dimensional systolic cross-sectional area comparison at three locations ($x/L=$0, 0.5, 1) in the MCA, for seven scenarios; baseline (no vasospasm), and six levels of vasospasm starting from $10\%$ lumen diameter reduction (mild vasospasm) to $60\%$ (severe vasospasm) diameter reduction in $10\%$ increments. \textbf{g)} Comparison of PINN predictions of CVS cross-sectional areas and the clinically utilized hemispheric ratio (HR).}
	\label{fig::fig4}
\end{figure}
\par 
\par 
For this purpose, we generated synthetic data by performing axisymmetric one-dimensional CFD simulations of pulsatile blood flow for the same CoW arterial tree with and without vessel stenosis. We used the outputs of the CFD simulations (velocity, area, and pressure in the entire CoW vasculature) for pre-spasm baseline as well as six levels of vessel spasm  from $10\%$ (mild vasospasm) to $60\%$ (severe vasospasm) diameter reduction \cite{li2019review,melis2019improved}, which are equivalent to $19\%$ to $84\%$ cross-sectional area reduction, respectively. For all the cases, we used the same inlet and outlet boundary conditions as explained in section \ref{sec::Results_hemodynamic}, which although not realistic, still exhibits the capability of our framework to detect the vasospasm by computing the local vessel cross-sectional areas employing only the updated velocity data. Moreover, since the PINN model is agnostic with respect to the boundary conditions, any physiologically change in the boundary conditions due to the vessel stenosis would reflect in the velocity TCD measurements and therefore the PINN model learns how to update areas accordingly.
% The  levels of vasospasm narrowing were based on a recent work for which two fully trained interventional neuroradiologists from the University Hospital of Tours in France were consulted \cite{melis2019improved}. 
We imposed the vessel narrowing in the M1 segment of the left MCA of the CoW network since it is the most commonly occurring region for CVS \cite{neulen2019volumetric,melis2019improved}. Furthermore, as previously reported, we assume that narrowing is uniform across the lumen in the affected branch, with no changes in vessel shape \cite{melis2019improved}. We then solve an inverse problem by retraining our PINN models to estimate and update the vessel cross-sectional area maps in the entire CoW after the vessel stenosis and compared its predictions against the CFD ground-truth (GT) simulations in terms of vessel cross-sectional area and hemodynamic variables. Our main assumption here is that the blood pressure does not substantially alter after vessel stenosis and thus we can utilize the pressure distributions in the vasculature before CVS to recompute the vessel cross-sectional areas after CVS (see appendix \ref{sec::appendix_A} \textbf{Fig. S3}). %(see supplementary \textbf{Fig. \ref{fig::Suppl.Fig3}}). 
%Briefly, based on the one-dimensional blood flow modeling at the baseline condition (no vasospasm), we first train a series of PINN models for each segment in the CoW to predict the velocity, area, and blood pressure in the whole vasculature. Next, with the information of the updated blood flow velocities after CVS at a certain number of spatial positions along with the spatio-temporal pressure distributions, we train another set of neural networks for each segment after CVS such that their predictions in pressure and velocities fit the data, while their predictions are enforced to follow the governing equations of one-dimensional pulsatile blood flow model \cite{sherwin2003one}. Note that after vasospasm the equilibrium cross-sectional area of each vessel ($A_{0}$) is updated during training (See section \ref{sec:Methods_CVS} for more details). 
We took the CFD velocity time series at the same nine points as shown in \textbf{Fig. \ref{fig::fig2}a} to mimic the clinical TCD measurements, which we used in order to train the spastic network (\textbf{Fig. \ref{fig::fig4}a}). We observed that the blood flow velocities in the MCA artery are significantly altered compared to other vessels (red box), given its proximity to the location of stenosis. On the other hand, the blood velocities in the vessels that are sufficiently away from the stenosed MCA (blue and purple boxes) are less affected. 
\par 
To further demonstrate the performance of the PINN model, cross-sectional areas and blood velocities at systolic phase after vasospasm, recomputed by PINN, are projected on the vessel centerlines (\textbf{Fig. \ref{fig::fig4}b-c}). We found good agreement between the predictions of PINN and the ground-truth data provided by the CFD simulations, indicating the capability of the presented approach in updating the local vascular cross-sectional areas and hemodynamics in the entire vasculature. In addition, the area and velocity space-time maps for the affected vessel (M1 segment of the left MCA) at $30\%$ reduction in diameter are compared between the PINN predictions and ground truth (\textbf{Fig. \ref{fig::fig4}d-e}). To better visualize the two maps, the cross-sectional areas are non-dimensionalized by subtracting and normalizing by the equilibrium cross-sectional areas $A_{0}$ at each spatial position $x$, \textit{i.e.}, $\hat{A} \left(x,t\right) = \left(A\left(x,t\right) - A_{0}\left(x\right)\right)/ A_{0}\left(x\right)$. 
We further compared the non-dimensionalized cross-sectional area $\hat{A}$ in the affected MCA vessel at three positions $(x/L \in [0,0.5,1])$ corresponding to beginning, middle, and end points along the vessels longitudinal axis and at peak systolic phase (\textbf{Fig. \ref{fig::fig4}f}). The data points are tightly distributed along the 45-degree line (shown by red dashed line), indicating good agreement between the PINN and the GT. We presume that the pressure distributions remain unchanged from baseline condition. Nonetheless, they deviate from baseline conditions for severe vasospasm, \textit{i.e.,} $60\%$ lumen diameter reduction after vessel stenosis (see appendix \ref{sec::appendix_A} \textbf{Fig. S3}). %(see supplementary \textbf{Fig. \ref{fig::Suppl.Fig3}}). 
Consequently, the predictions of PINN are prone to inaccuracies in the case of severe vasospasm.
% It is worth mentioning that all of this is accomplished merely based on updated velocity time-series at the nine locations on the CoW, similar to the clinical TCD measurements, and does not require changes to the clinical workflow (see section \ref{sec:Methods_CVS} for more details on the approach). \hb{Too enthusiastic at the risk of over-claiming. } 
To further demonstrate the clinical feasibility of our  approach, we compare the cross-sectional area predictions to the hemispheric ratio (HR) also known as ``Lindegaard ratio" \cite{lindegaard1989cerebral} for baseline condition and all stages of vasospasm (\textbf{Fig. \ref{fig::fig4}g}). The HR is the ratio between mean flow velocities through the MCA and ipsilateral ICA \cite{lindegaard1989cerebral}. This ratio was recommended to improve the accuracy of TCD measurements for vasospasm detection at the bedside and ratios larger than three indicate anterior vasospasm \cite{lindegaard1989cerebral}. This ratio has also been well-captured by the PINN model as \textbf{Fig. \ref{fig::fig4}g} demonstrates the HR values are larger than 3 (corresponding to $10\%$ reduction in vessel lumen diameter) for all vasospasm cases. As expected, the vessel's local diameter shrinkage due to vasospasm leads to an increase in blood flow velocities and higher HR values. 

% ########################################
%#################################################################################
%#################################################################################
\section{Discussion} \label{sec::Discussion}
Developing a technology that is able to generate rapid and reliable estimates for brain hemodynamics as well as local vessel diameters with high spatiotemporal resolution can significantly improve diagnosis and treatment of various cerebrovascular diseases. Currently, TCD ultrasound is the most common technique that is utilized in clinical practice to approximate hemodynamics albeit at only a few cross-sections accessible through the cranial acoustic windows, which might not be sufficient to gather comprehensive insight for clinical diagnosis. In addition, CTA and ToF scans, although highly reliable in terms of extracting vascular geometrical features, are limited for repeated measurements. 
In this work, we put forth a deep learning framework that augments the sparse clinical data, including baseline admission angiograms and repeated daily TCD measurements, with simplified physical laws of fluid flow in compliant arteries to recover brain hemodynamics and changes in local vessel cross-sectional areas in the entire brain vasculature. We developed a physics-based neural network model called ``ASPINN" in order for use in cerebrovascular hemodynamics with commonly collected clinical data.   
% \textit{i.e.} real-time, non-invasive TCD ultrasound and ToF or CTA data and experiment the model on the more complex arterial topologies \textit{i.e,} cerebral vasculature. 
We validated our approach by comparing the predicted blood velocity time series at several locations in the CoW network against \textit{in-vivo} velocity data acquired from 4D flow MRI. Our framework can lead to a significant paradigm shift in quantitative approaches towards clinical sciences, wherein by augmenting the highly accurate (yet sparse) clinical measurements with the more easily available computational models in order to increase the spatiotemporal
resolution while maintaining the same level of accuracy of the clinical measurements.
\par 
We further showed the clinical significance of our developed deep learning model by utilizing it to diagnose CVS, which is one of the most devastating cerebrovascular diseases by solving an inverse problem. We demonstrated this clinical value by synthesizing blood flow through the CoW for the baseline condition (no vasospasm) and after vasospasm via CFD simulations. We found that the PINNs can accurately estimate the vessel cross-sectional areas and brain hemodynamics maps in the entire vascular domain using only the baseline angiograms (representing ToF MRI or CTA), updated velocities (representing TCD measurements) at the same nine locations as shown in \textbf{Fig. \ref{fig::fig2}}, and pressure distributions given by the PINNs from the baseline condition. 
Although we generated synthetic data to show the clinical value of our developed tool, the same tool can be applied in the clinical case where we do have access to the admission CTA and daily TCD data. Eventually, the estimated vessel local diameters resulting from the PINN predictions can be compared against a secondary CTA, which is usually acquired after vessel spasm in the current clinical workflow. %We posit that our algorithm should provide higher sensitivity and specificity compared to TCD ultrasound currently used for bedside CVS diagnosis. This is due to the fact that via the PINN algorithm we update the primary CVS diagnosis variable, \textit{i.e.,} vessel local diameter in the entire vascular tree daily and at the bedside rather than blood flow velocities as a surrogate for vascular diameter. Hence, the diagnosis can be significantly improved and on par with CTA, which is the gold standard technique for CVS diagnosis. %Further, the diagnosis can be performed at the bedside and thus eliminates the CTA limitations as mentioned in section \ref{sec::intro}.
\par 
%By simulating blood flow using a purely physics-based CFD model \cite{sherwin2003one}, 
We found that although the CFD results qualitatively agree with \textit{in-vivo} clinical measurements, they tend to have substantially larger deviations in the distribution of the hemodynamic variables compared to PINN results. 
% Simulating the blood flow through the CoW network with conventional purely physics-based computational models \cite{sherwin2003one} shed light on the physics deficiencies of these models and the critical role of imposing physiologically correct outlet boundary conditions, which are generally unknown. We found substantial deviations in the distribution of the brain hemodynamic variables from the \textit{in-vivo} clinical data and the neural network predictions. 
% Although these simplified CFD models qualitatively agree with \textit{in-vivo} clinical data for cerebral arteries \cite{huang20181d}, here, through performing the quantitative comparison, we do actually observe a large discrepancy between the model predictions and the clinically acquired data. 
Conversely, the PINN approach trained on the clinical data, despite the physics-deficiencies of the same underlying axisymmetric model, recovers the brain hemodynamics with higher accuracy. Furthermore, the deep learning model does not suffer from the major limitations imposed by the pure computational models such as precise prescription of boundary conditions, hefty mesh generation, or even constitutive laws \cite{raissi2019physics,raissi2020hidden,kissas2020machine}.  
% \par 
% Traditional CFD tools require an exact prescription of boundary conditions. 
In hemodynamics simulations, in particular, inaccurate prescription of inflow/outflow boundary conditions can severely degrade the simulation predictions \cite{madhavan2018effect,rajabzadeh2020inter}. Although blood velocity measurements via TCD are available, the spatially sparse measurement points usually lie in the interior of the domain of interest and not at the boundaries. Incorporating these measurements into the CFD code is challenging because: (\textit{i}) one has to solve an inverse problem to infer the boundary conditions from these measurements, which can be cast as an optimization problem and it is in general  costly to solve, and often becomes ill-conditioned or under-determined in the absence of sufficient measurements, (\textit{ii}) the TCD measurements are  noisy, precluding their utility in numerical techniques that demand smooth profiles for differentiation. On the other hand, our PINN approach does not require the exact prescription of the boundary conditions and they compute derivatives and solve inverse problems using  noisy data. 
% \par
The results from our PINN model will further improve the axisymmetric CFD simulations of pulsatile blood flow in deformable walls. Indeed, it appears that the purely physics-based CFD models can predict the general trend of hemodynamic variables rapidly. However, they fail to capture the velocity peak and mean values and induce a time delay in a cardiac cycle. This is due to the combined effects of physics deficiencies and uncertainties in outlet boundary condition subscription. The former is the result of model simplifications and the latter is due to the lack of knowledge in internal blood pressure distribution in efferent vessels. Conversely, our PINN model, which is trained to both fit the clinical data and satisfy the physical laws of unsteady blood flow, can successfully capture blood velocity peak and mean values as well as its spatiotemporal distribution. %Furthermore, it predicts the pressure distribution in the entire vasculature including the efferent vessels that can be subsequently employed to compute physiologically correct and subject-specific outlet boundary conditions. 
Therefore, our findings in this paper may be used as a benchmark for refining the CFD models to better capture brain hemodynamics.
\par 
The current study has a number of limitations that should be considered when interpreting the results.
In this work, for the vessel wall compliance, we resorted to the simplest constitutive law that correlates pressure and area, \textit{i.e.,} the elastic tube law, neglecting the effects of the vessel wall viscoelasticity on pressure distribution \cite{alastruey2011pulse}. In addition, the axisymmetric model that we have adopted neglects 3D effects such as vessel tortuosity and curvature which are expected to play a major role in predicting hemodynamic factors for cerebral arteries. Although incorporating more sophisticated models into the PINN framework could lead to more accurate results \cite{kissas2020machine,formaggia2010cardiovascular,lamponi2004one}, we found that, despite using the simple axisymmetric model, our approach is capable of generating hemodynamic results on par with \textit{in-vivo} data in terms of accuracy, while ensuring much less computationally-intensive and faster results. 
Furthermore, in order to train the PINNs properly, the spatial position of the velocity training data in the arterial tree should be specified. Since velocity data is obtained via TCD, that information is elusive. In this work, in order to determine the location of TCD velocity measurements, we used the 4D flow MRI measurements, which is usually unavailable in clinical measurements. Having further studied the effect of precise knowledge of the location, we found that for the vessels with almost constant cross-sectional area in space, the model predictions are fairly robust, but for the vessels with varying cross-sectional areas, furture studies should be performed to test the robustness of the PINNs. 
We also assumed that the vessel areas resulting from segmentation of the ToF scan is at the equilibrium position due to the lack of information on the exact time of image acquisition in a cardiac cycle. However, we found that this assumption does not significantly affect the predictions of PINNs as wall displacements of intracranial arteries in a cardiac cycle are small (less than $3\%$) due to the high vessel stiffness \cite{alastruey2007modelling}. 
Additionally, as the TCD angle of insonation is unknown, the TCD velocity values are not corrected by the cosine of this angle, which is common clinical practice. We presume that this angle for major arteries in the CoW \textit{e.g.} MCA is small \cite{grolimund1987evaluation}, so the measured values correspond well to true velocity. 
Finally, to update the vessel local diameters and detect the vasospasm, we assumed that the pressure distribution in the affected vessel does not change significantly as compared to the baseline condition. However, this is not the case for all the CVS cases, especially for the severe vasospasm, \textit{i.e.,} $60\%$ lumen diameter reduction. %(see supplementary \textbf{figure \ref{fig::Suppl.Fig3}}). 
Similarly, our algorithm relies on an assumption that the vessel shape is not greatly affected by the vasospasm and the vessel diameter is reduced uniformly across the lumen. Although previous researchers made the same assumption in their simulations \cite{melis2019improved} or observed the same pattern in their experiments \cite{sloan1989sensitivity}, uniform diameter reduction may not always occur in the SAH patients. Consequently, the pressure distribution will alter after vessel spasm as compared to the baseline conditions and cannot be used to retrain the PINNs and update the area maps. Hence, the PINN framework with its current format may generate inaccurate vessel cross-sectional areas and brain hemodynamic parameters due to the lack of sufficient training data for solving an inverse problem.
\par 
In summary, this study opens an avenue in potentially addressing many clinical problems regarding cerebrovascular diseases and introduces a new technology that can be readily transferred to the clinical workflow and help clinicians to diagnose and make better decisions for the subsequent treatment. Moreover, this study sheds light on the physics deficiencies of conventional reduced-order computational models of pulsatile blood flow and highlights the importance of accurate boundary condition subscription in predicting brain hemodynamics. We demonstrated that by combining the most simple reduced-order computational model with the \textit{in-vivo} clinical data through deep learning algorithms, we can recover the hemodynamics and wall displacements of brain arteries with an accuracy comparable to clinical data, albeit with higher spatiotemporal resolution. The pressure distribution predicted by the PINNs can be employed to discover the physiologically correct outlet boundary conditions and hence eliminates one of the sources of uncertainties in conventional CFD models.    

\appendix
\beginsupplement
\subsection{MRI data analysis} \label{sec::appendix_A}
We acquired the arterial geometry of the CoW from a healthy male volunteer (age = 30 years, weight = 94 kg, height = 185 cm) by measuring 200 slices
of $0.25mm$ thickness with a 3D-TOF MRI sequence. The in-plane resolution
was $0.4mm \times 0.4mm$, the repetition time (TR) and the echo time (TE) were $18 ms$ and $3.57 ms$, respectively, and the flip angle was 15\textdegree. %\ang{15}.
In the same scanning session, a prospectively ECG-gated 4D flow MRI scan was performed using a 3.0 Tesla MRI scanner (Skyra, Siemens Healthcare, Erlangen, Germany) and a 32-channel head coil. Scanning parameters were as follows: flip angle = 7\textdegree, repetition time/echo time=44.56/2.78 ms, bandwidth=445 Hz/pixel, velocity encoding (VENC)=90 cm/s, voxel size =1.5x1.26x2 $mm^3$, temporal resolution = 41.36 ms over 14 cardiac phases.   
\par 
From the 3D-TOF slices, the 3D geometry was segmented with a model construction pipeline of SimVascular. The threshold for the basic segmentation was set to one-fourth of the maximum signal intensity value obtained in the measurements, which showed the best results for segmenting arterial structures. Everything above this value was considered to be an artery. Additional segmentation was performed manually adding or removing voxels from the geometry based on anatomical knowledge.
\par 
Before extracting velocity from 4D flow MRI phase difference images, we perform image pre-processing including noise-masking, eddy-currents correction as well as phase unwrapping following the method explained in. Noise masking was performed by thresh-holding of the signal intensity in the magnitude data to exclude regions with low signal intensity. Eddy current correction consisted of three steps: 1. separation of static regions from blood flow, 2. fit a plane with least-squares method to the static regions from the last time frame (late diastole), 3. Subtraction of the fitted plane to the MRI data in every time frame. Finally, we correct the velocity aliasing effect, \textit{i.e,} if adjacent pixels velocities in temporal or slice direction differ by more than VENC. If aliasing occurs, we unwrap the phase difference images based on the following equation:
\begin{equation} \label{eq:MRI_Velocity_aliasing}
V \left(x,y,z,t\right) = V \left(x,y,z,t\right) - 2\sign\left(V\left(x,y,z,t\right)\right)\VENC.
\end{equation}
Once, all the pre-processing steps are done and the image, we remap the phase difference images in the range of $-\pi \leq \Delta \phi \leq \pi$ and compute the blood velocity via:
\begin{equation}\label{eq:MRI_Velocity}
V \left(x,y,z,t\right) = \frac{\Delta \phi \left(x,y,z,t\right)}{\pi} \VENC.
\end{equation}
All the medical image processing steps are performed using an in-house MATLAB code.
%\subsection{Figures}
\begin{figure}[h!]
	\centering
	\includegraphics[trim={0cm 0cm 0cm 0cm},width=1\linewidth]{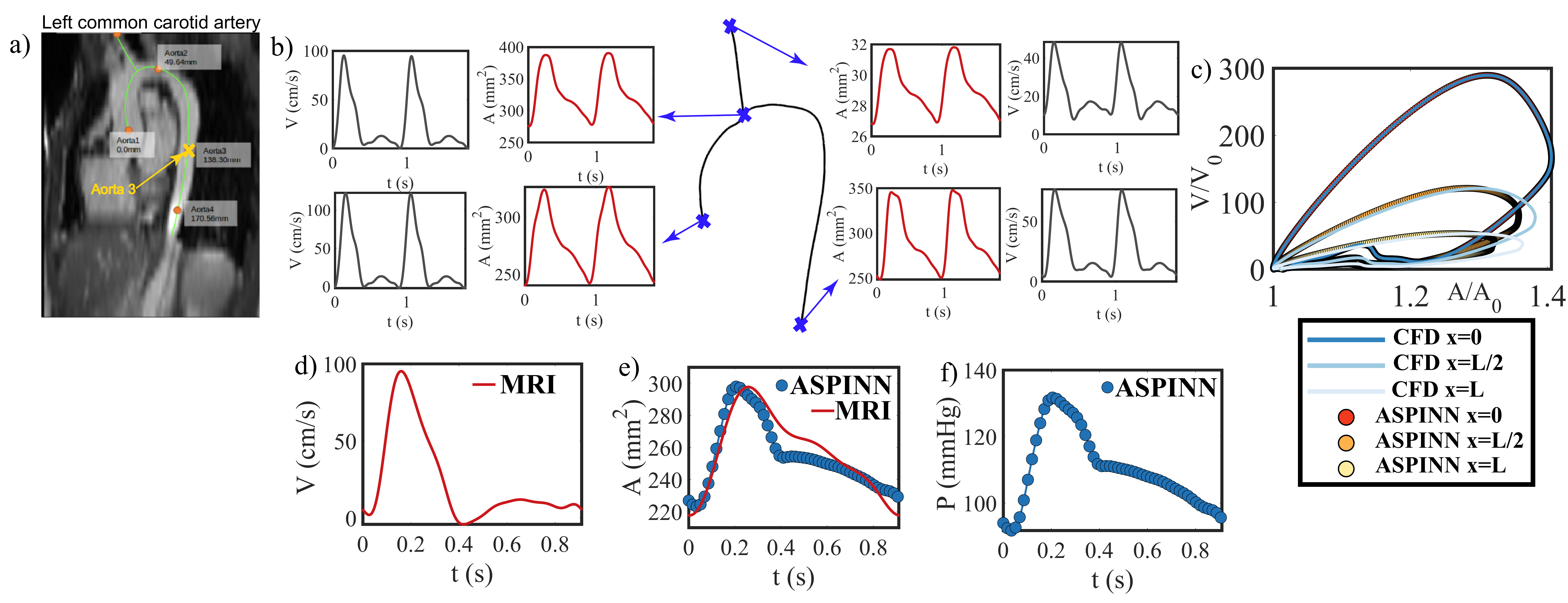} 	
	%	\vspace{-30ex}
	\caption{Validation of the developed area surrogate physics-informed neural networks (ASPINN) that approximate the subject-specific velocity-area correlation on flow through the aorta/common carotid bifurcation of a healthy human volunteer. a) 4D flow MRI magnitude image along with the positions of velocity and area measurements in the aorta/carotid bifurcation of a healthy subject. Area time-series measurement at Aorta 3 (yellow cross-marker) is used to validate the ASPINN predictions. b) Vessel centerlines and the velocity and area time evolution in two cardiac cycle resulting from one-dimensional computational modeling at four spatial positions in the arterial tree (marked by blue cross-markers) which include inlet of the aorta (Aorta 1), aorta and left common carotid bifurcation, outlet of the left common carotid artery, and outlet of aorta (Aorta 4). c) Comparison of velocity-area correlations at three spatial positions along the vessel that connects Aorta 2 and Aorta 3 between the ASPINN predictions and one-dimensional computational fluid dynamics (CFD) modeling of the pulsatile blood flow. d) Velocity measurement from 4D flow MRI at Aorta 3 (yellow cross-marker in panel a)) over one pulsatile cycle. e) Comparison of area time evolution resulting from the ASPINN prediction and the 4D flow MRI measurements at Aorta 3. f) Pressure time-series at Aorta 3 resulting from the ASPINN prediction. Clearly, the optimized ASPINN that is trained to recover subject-specific velocity-area correlation, can estimate the wall displacement with high accuracy. Further, while we cannot validate the pressure distribution due to the lack of clinical data, the range of predicted pressure by the ASPINN is physiological.}
	\label{fig::Suppl.Fig1}
\end{figure}

\begin{figure}[h!]
	\centering
	\includegraphics[trim={0cm 0cm 0cm 0cm},width=1\linewidth]{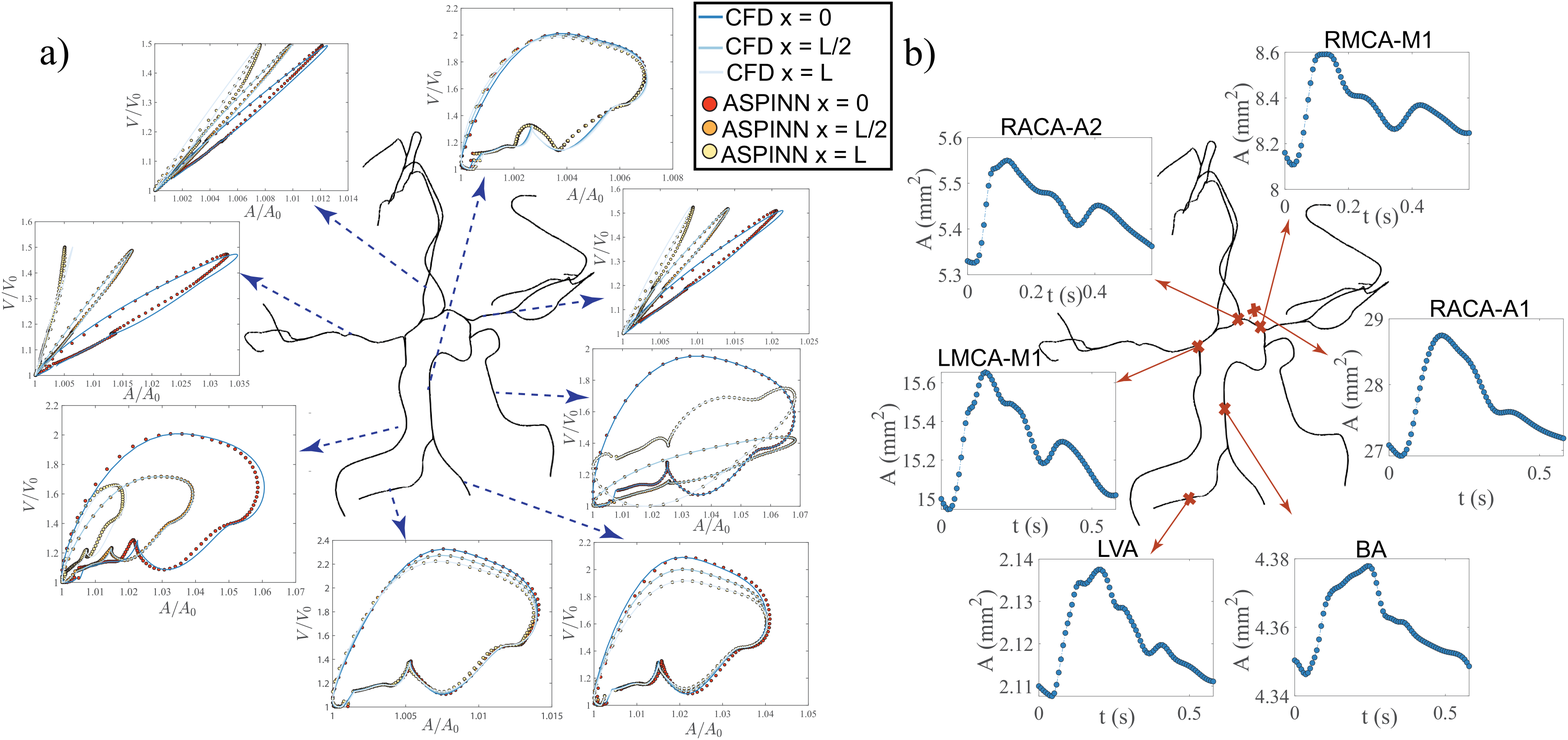} 	
	%	\vspace{-30ex}
	\caption{a) Area surrogate physics-informed neural networks (ASPINN) predictions of velocity-area correlations in different segments and at three spatial positions in each segment of the circle of Willis arterial network. The predicted correlations using the ASPINN model are also compared against one-dimensional CFD modeling. The correlations are shown at the beginning, middle and endpoints of the left and right vertebral arteries (LVA, RVA), left and right sides of internal carotid arteries (LICA, RICA), basilar artery (BA), left and right sides of M1-segment of middle cerebral arteries (LMCA-M1, RMCA-M1), and A2-segment of the left anterior cerebral artery (LACA-A2). b) ASPINN model predictions of area time-series in vessels of the CoW network (marked by orange cross-markers) that we have access to the Transcranial Doppler (TCD) ultrasound velocity measurements. These generated data will be subsequently used as training data in the PINN model to predict the brain hemodynamics in the entire vasculature. Note that, in total, we have the velocity measurements in nine segments of the vasculature. However, here we show the area predictions of the ASPINN model at six vessels for the sake of clarity. }
	\label{fig::Suppl.Fig2}
\end{figure}

\begin{figure}[h!]
	\centering
	\includegraphics[trim={0cm 0cm 0cm 0cm},width=1\linewidth]{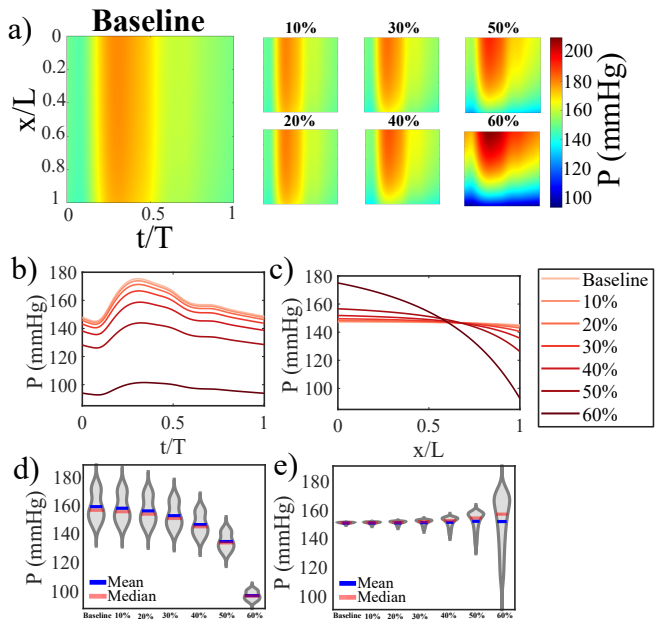}
	%	\vspace{-30ex}
	\caption{Pressure distributions in the MI-segment of the left MCA of the CoW arterial network at baseline condition (no vasospasm) and six stages of vasospasm \textit{i.e,} $10\%$ (mild vasospasm), $20\%$, $30\%$, $40\%$, $50\%$, and $60\%$ (severe vasospasm) of lumen diameter reduction. In all of the vasospasm cases, narrowing is uniform across the lumen, with no alter in vessel shape. a) Comparison of blood pressure colormaps in the $x-t$ plane between baseline condition and the vasospasm cases. b) Comparison of pressure time-series at the vessel endpoint \textit{i.e,} the bifurcation point of M1-segment and M2-segments (daughter branches) of the left MCA between baseline condition and vasospasm cases. c) Comparison of blood pressure distribution at diastolic phase along the vessel between baseline condition and vasospasm cases. d) Probability density functions (PDFs) of blood pressure during one cardiac cycle at the vessel endpoint for the baseline and six stages of vasospasm cases along with the mean and median values. e) PDFs of the blood pressure along the vessel length at the diastolic phase for the baseline and vasospasm cases. The mean and median values are indicated.}
	\label{fig::Suppl.Fig3}
\end{figure}

\begin{figure}[h!]
	\centering
	\includegraphics[trim={0cm 0cm 0cm 0cm},width=1\linewidth]{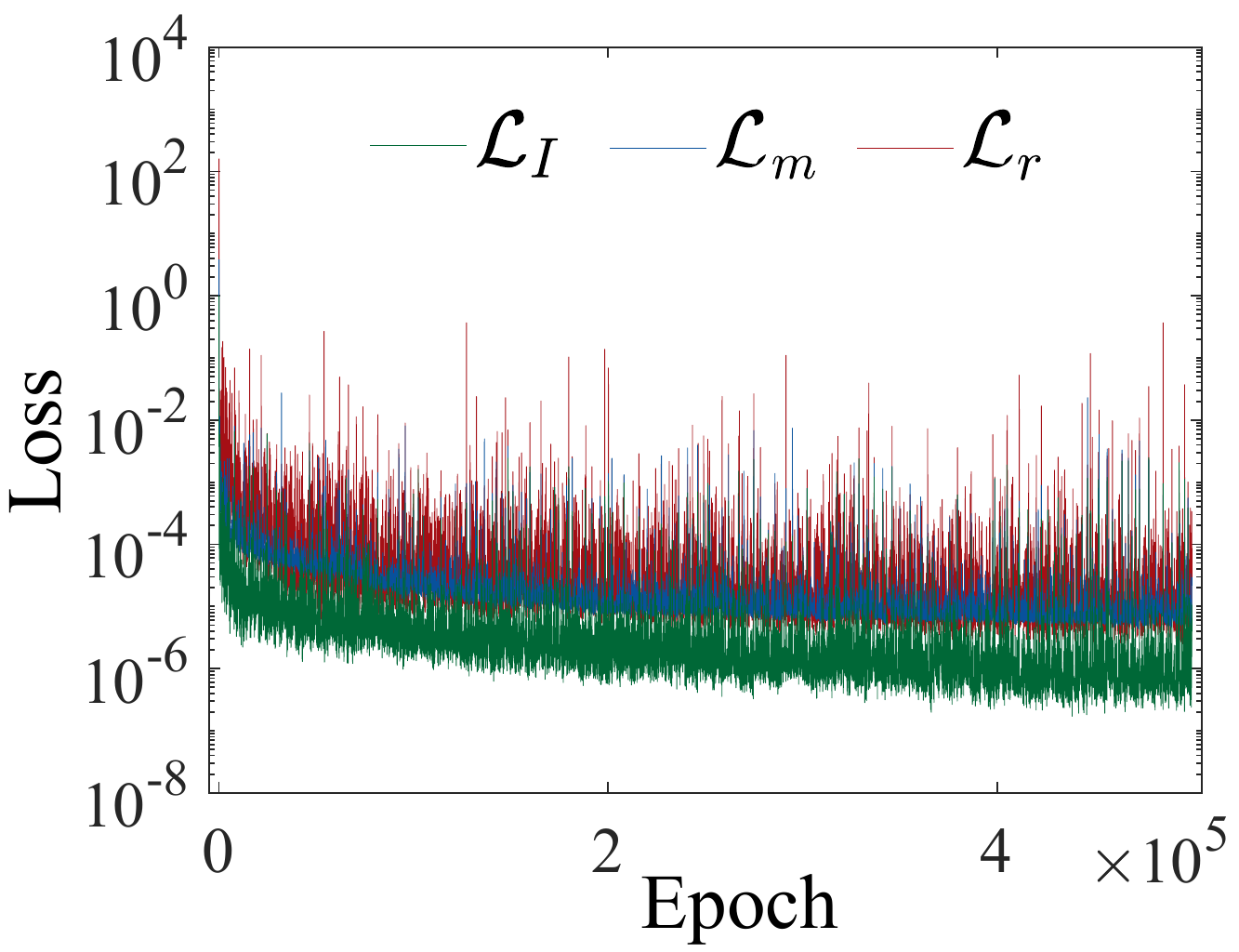} 	
	%	\vspace{-30ex}
	\caption{Flow through the CoW network of a healthy human subject: Values of the loss functions versus number of stochastic gradient descent epochs during the training of the PINN model. The solid green line corresponds to the interface loss $\mathcal{L}_{I}$. The solid blue line denotes the measurement loss of velocity measurements, and the area generated by the area surrogate physics-informed neural networks (ASPINN) model $\mathcal{L}_{m}$. The solid red line represents the residual loss $\mathcal{L}_{r}$.}
	\label{fig::Suppl.Fig4}
\end{figure}

\end{document}